\documentclass[preprint,superscriptaddress,showpacs,amsmath,amssymb,aps,pre,floatfix]{revtex4-1}
\usepackage{color}
\usepackage{fullpage}
\usepackage{graphicx}
\usepackage{titlesec}
\usepackage{titletoc}
\usepackage{stmaryrd}
\usepackage{hyperref}
\usepackage{hhline}
\usepackage{bm}
\usepackage{bbm}
\usepackage{mathbbol}
\usepackage{booktabs}
\usepackage{setspace}
\usepackage{csvsimple}
\usepackage[normalem]{ulem}
\usepackage{longtable}
\usepackage{dcolumn}
\usepackage{float}
\usepackage[resetlabels,labeled]{multibib}

\newcites{New}{The other list}

\AtBeginDocument{
\heavyrulewidth=.08em
\lightrulewidth=.05em
\cmidrulewidth=.03em
\belowrulesep=.65ex
\belowbottomsep=0pt
\aboverulesep=.4ex
\abovetopsep=0pt
\cmidrulesep=\doublerulesep
\cmidrulekern=.5em
\defaultaddspace=.5em
}

\begin{document}
\title{Interplay between intra-urban population density and mobility in determining the spread of epidemics.}
\author{Surendra Hazarie}
\thanks{These two authors contributed equally to this work.}
\affiliation{Department of Physics \& Astronomy, University of Rochester, Rochester, NY, 14627, USA.}
\author{David Soriano-Pa\~nos}
\thanks{These two authors contributed equally to this work.}
\affiliation{GOTHAM Lab -- Department of Condensed Matter Physics and Institute for Biocomputation and Physics of Complex Systems (BIFI), University of Zaragoza, E-50009 Zaragoza, Spain.}
\author{Alex Arenas}
\affiliation{Departament d'Enginyeria Inform\`atica i Matem\`atiques, Universitat Rovira i Virgili, E-43007 Tarragona, Spain.}
\author{Jes\'us G\'omez-Garde\~nes}
\thanks{gardenes@unizar.es}
\affiliation{GOTHAM Lab -- Department of Condensed Matter Physics and Institute for Biocomputation and Physics of Complex Systems (BIFI), University of Zaragoza, E-50009 Zaragoza, Spain.}
\affiliation{Center for Computational Social Science (CCSS), Kobe University, Kobe 657-8501, Japan.}
\author{Gourab Ghoshal}
\thanks{gghoshal@pas.rochester.edu}
\affiliation{Department of Physics \& Astronomy, University of Rochester, Rochester, NY, 14627, USA.}
\affiliation{Department of Computer Science, University of Rochester, Rochester, NY, 14627, USA.}

\begin{abstract}
\singlespacing
In this work, we address the connection between population density centers in urban areas, and 
the nature of human flows between such centers, in shaping the vulnerability to the onset of contagious diseases. A study of 163 cities, chosen from four different continents reveals a universal trend, whereby the risk induced by human mobility increases in those cities where mobility flows are predominantly between high population density centers. We apply our formalism to the spread of SARS-COV-2 in the United States, providing a plausible explanation for the observed heterogeneity in the spreading process across cities. Armed with this insight, we propose realistic mitigation strategies (less severe than lockdowns), based on modifying the mobility in cities. Our results suggest that an optimal control strategy involves an asymmetric policy that restricts flows entering the most vulnerable areas but allowing residents to continue their usual mobility patterns.
\end{abstract}

\maketitle

\section{Introduction}
During the last century, humankind has rapidly evolved into an interconnected society driven by the existence of a vast mobility network connecting different areas around the globe. In particular, the striking growth experienced by the international mobility network \cite{aviationgrowth} has helped to bridge socio-cultural \cite{varghese2008globalization,filatotchev2011knowledge,williams2005human} and economic gaps \cite{boubtane2016immigration}. Accompanying this is the phenomenon of urbanization, whereby a majority of the world's population reside in densely packed urban centers, with the trend only accelerating~\cite{urbanization, Le2012}, given the socioeconomic advantages that cities afford~\cite{bettancourt2013, Pan_2013}.   Allayed against these benefits, the increase in mobility has removed the main bottleneck limiting the spatial diffusion of confined epidemic outbreaks. Once the disease spreads to different regions, it takes advantage of the high population density and infrastructure networks in cities~\cite{Kirkley_2018} to rapidly spread through the local population. As a consequence, over the past few years, several contagious disease outbreaks have emerged, notable  among them H1N1 in 2009~\cite{Van-Kerkhove:2011nm}, Ebola in 2014 \cite{vespiebola2014}, ZIKV in 2015 \cite{zika2017} and of course more recently SARS-COV-2 \cite{Estrada2020,world2020coronavirus}. Indeed, the frequency with which these pandemics occur is troublingly increasing~\cite{SH_2016}.

Despite the different nature of the pathogens, their spreading, both globally and locally, is primarily explained by the sequential combination of case importation from contagion sources, followed by local community transmission converting initially unaffected regions into new endemic areas. Different flavors of mobility play a role in each process: on the one hand, long-haul flights \cite{bowen2006airline} are usually the drivers facilitating the entry of pathogens into a given country, to the extent that the airport mobility network has proved to be a reliable proxy to determine pathogen arrival times \cite{Brockmann2013}, international infection routes \cite{airport2007}, and to complement phylogeographic inference of emerging pathogens such as SARS-CoV-2 \cite{Lamey2020}.  On the other hand, once index cases are found within a given region, a complex combination of the local mobility \cite{Barbosa2018} and socio-economic features of the population \cite{bonaccorsi2020economic,ahmed2020inequality} determines the speed of epidemic spread and the extent of its outbreak. 

Quantifying the impact of local mobility on the global diffusion of a pandemic constitutes a challenging task. In this sense, several examples addressing the impact of daily recurrent mobility patterns on the spread of contagious diseases can be found in the literature \cite{Pastor-Satorras2015,Belik2011,Balcan2011,Gomez-Gardenes2018,Soriano-Panos2018,Australiarecurrent,granell2018epidemic,Soriano-Panos2020}.
The majority of these, however, are theoretical frameworks analyzing the features of synthetic mobility networks, and the influence of total volume of travelers on the course of the epidemic. Nonetheless, recent advances made in data-gathering techniques allow for obtaining accurate representations of daily urban rhythms constructed from mobile phone traces \cite{Tizzoni2014}, geolocalized tweets \cite{Mazzoli2020}, Location Based Social Networks~\cite{Barbosa2015}, or extensive census surveys. These data sets enable the extension of the theoretical machinery to address real epidemic scenarios. Indeed, recurrent mobility patterns have been already useful for identifying the most exposed areas in some epidemic scenarios \cite{soriano2020vector} as well as reproducing the infection routes of H1N1 influenza \cite{Influenza2009}, Malaria \cite{Wesolowski267}, and more recently SARS-COV-2 \cite{Arenas2020,Costa2020,USACOVID,ItalyCOVID}. 

While much attention has been spent on reconstructing past infections, or epidemic forecasting in the case of extant pandemics, an important question that immediately arises, is  what makes regions---in particular urban agglomerations where most people reside---vulnerable to the spread of pathogens in the first place? While factors such as population density, levels of healthcare, quality of infrastructure and socioeconomic disparities play a major role~\cite{Alirol:2011uj}, vulnerability to spread is a complex interplay between these features that is, in general, difficult to disentangle. For instance, the role of population density is an open question with evidence both for and against its influence on epidemic spreading~\cite{Kraemer:2015yw,Li:2018ce}. Indeed, merely the density of contacts, while relevant at a neighborhood  level, is not enough to explain the mechanisms of spread; one would also need to consider the mobility network of flows that govern the exchange of people between the regions. In such a setting, the spatial distribution of the population densities and the strength of interaction between the regions become especially relevant. 

In other words it is reasonable to assume that the morphology of the city in terms of how its residents are distributed and how they navigate the city plays a crucial role in their susceptibility to pandemics. Indeed, recent studies have shown that the spatial patterns of how residents utilize transportation infrastructure is a strong indicator of that regions' levels of social inclusion, quality of infrastructure and wealth creation~\cite{Lee2017}. In~\cite{Bassolas2019} the authors propose a measure of a city's dynamical organization based on mobility hotspots~\cite{Louail2014} to classify them in a spectrum between compact-hierarchical and  sprawled layouts. The extent to which cities are compact or sprawled serve as a low-dimensional proxy for various urban indicators related to quality of life, health and pollution. 

In this work we connect the dots between the morphology of human activity in cities, in terms of its associated mobility flows and the distribution of resident populations, and its effect on shaping the transmission of infectious diseases and their associated epidemic outbreaks. We collect data from 163 cities across four continents, on their population density at the zip-code level, and intra-urban mobility flows for the first half of 2020. Using this we extract population density hotspots (i.e. those areas with the highest concentration of residents) and measure the extent to which flows between hotspots dominate the total flows in the city. To capture epidemic spreading, we generalize a MIR (Movement-Interaction-Return) epidemic model~\cite{Gomez-Gardenes2018} that captures the interplay between recurrent mobility flows and the distributions of resident populations.  We derive the epidemic threshold,  representing the minimum infectivity per contact required to instigate an epidemic outbreak, and connect it with the distribution of flows among population density hotspots. In particular we show that, despite their ostensible differences in terms of spatial layout, evolutionary history, or levels of infrastructure, all considered cities lie on a universal curve capturing an inverse relationship between the epidemic threshold and the extent to which mobility flows are localized between hotspots. 

The results suggests an increased susceptibility to epidemics as a function of flows being concentrated between high population centers. 
%
%
As a proof-of-concept, we analyze the current SARS-COV-2 pandemic by quantifying the epidemic growth from the initial infection curves as an empirical proxy for city vulnerability and plotting it against our calculated epidemic threshold. The empirical trends match our theoretical formalism where cities with mobility concentrated primarily between hotspots are more vulnerable as compared to those with more egalitarian mobility distributions. Based on this observation, we propose a realistic mitigation policy that, being much less severe than draconian lockdowns, lowers the susceptibility of cities that lie in the vulnerable spectrum. 

\section{Results}

\subsection{Data}

The population density for cities at zip-code resolution was collected from national census bureaus \cite{us_pop,aus_gov,zaf_gov} and from high resolution population density estimates recently published by Facebook \cite{fb_pop}. Each of these cities correspond to the largest in their respective countries in terms of population size.  From this data, we extract $H_k$ population density hotspots for each city $k$ (varying from city-to-city) by applying a non-parametric method based on the derivative of the Lorenz curve~\cite{Louail2014, Bassolas2019} (for details of the calculation see Supplementary Section S1). 

The mobility flows within each city are sourced from the Google SARS-COV-2 Aggregated Mobility Research Dataset, and contain anonymized flows mobility flows aggregated over users who have turned on the Location History setting, which is off by default. The flows are between cells of approximately 5 km$^2$ for the period ranging November 3rd 2019 to February 29th 2020. For the purposes of our analysis we only consider the period before any mobility mitigation measures were initiated as a response to the SARS-COV-2 pandemic (see Supplementary Section S2 for further details). The flows are encoded in a matrix  ${\bf T}$ whose elements $T_{ij}$ 
correspond to the population out-flows from  $i$ to $j$ and whose diagonal elements correspond to self-flows. For each city $k$, given $H_k$ hotspots, we calculate the hotspot concentration, $\kappa_k$, defined as the fraction of total flows in the city that occur only between hotspots thus,
\begin{equation}
    \kappa_k = \frac{\sum\limits_{i, j \epsilon H_k}{T_{ij}}}{\sum\limits_{i,j} T_{ij}}.
    \label{eq:kappa}
\end{equation}
This metric lies in the range $0 \leq \kappa_k \leq 1$, with the limiting cases corresponding to flows exclusively between hotspots or only between hotspot and non-hotspot areas. The list of cities for each country, the administrative unit, and the hotspot concentration is shown in Tab. S1. The results show a wide spectrum of values ($ 0.05 \leq \kappa_k \leq 0.79$) both within and between countries, indicating significant variability in cities in terms of the morphology of human flows between population centers. 

\begin{figure}[t!]
    \centering
    \includegraphics[width=1.02\columnwidth]{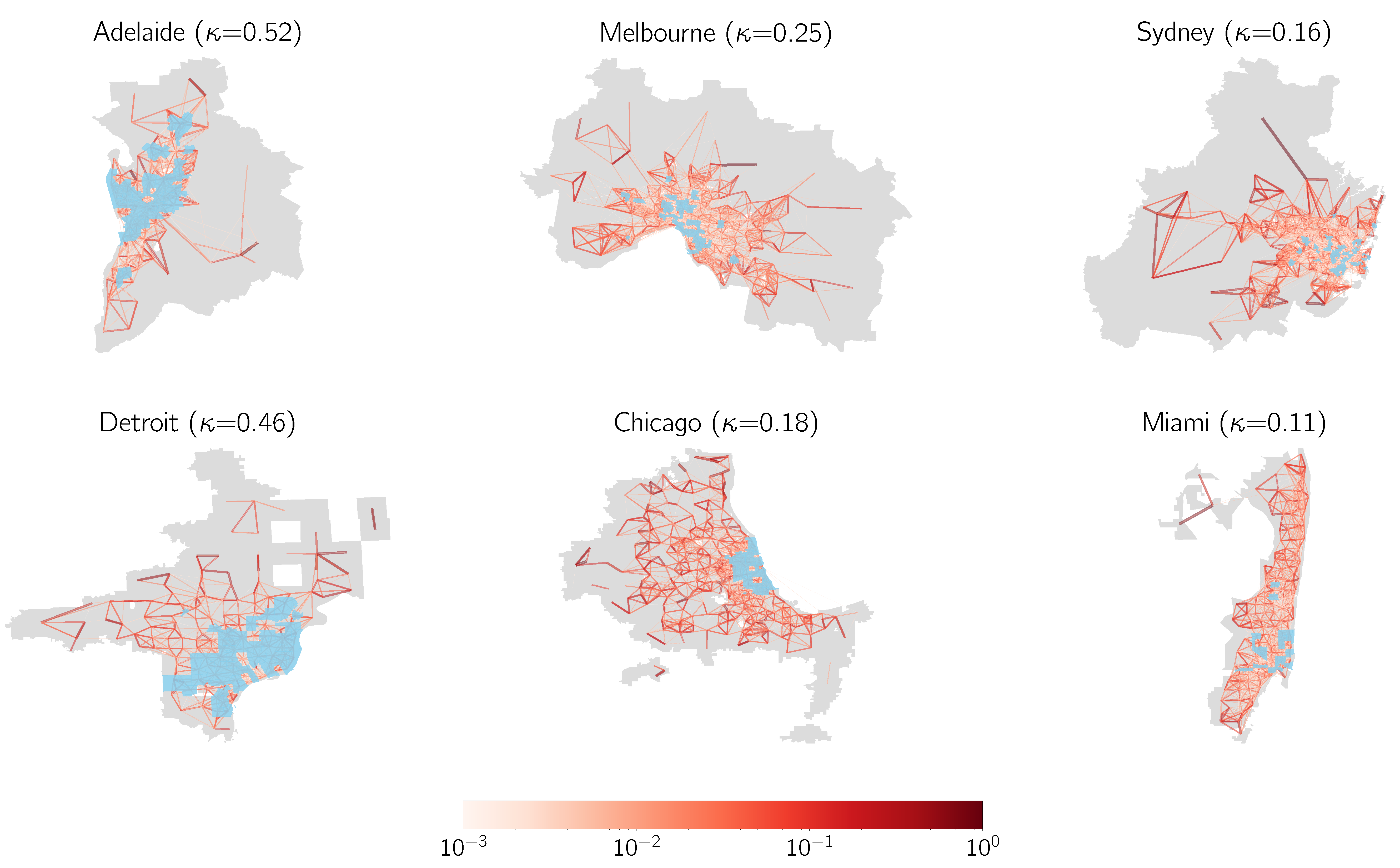}
    \caption{{\bf Spatial representation of the urban mobility network for selected cities} (upper panel) Autralia, (lower panel) United States. Areas marked blue correspond to population density hotspots. The cities are organized in descending order according to the extent to which mobility flows are concentrated between areas of high population density, $\kappa$ (Eq.\eqref{eq:kappa}). Line color encodes the number of inhabitants following each route normalized by the highest flow observed within each city.}
    \label{fig:scheme}
\end{figure}

In Fig.~\ref{fig:scheme}, we plot the spatial layout of the hotspots and the mobility network for six representative cities in Australia (upper panel) and the United States (lower panel). The cities are organized in descending order of $\kappa$, and it is apparent that in those cities with high $\kappa$, flows are mainly confined around hotspots (marked in red), whereas they are increasingly more distributed with decreasing $\kappa$. An additional feature is that hotspots are more spatially concentrated in those cities with lower $\kappa$ and dispersed in those with higher $\kappa$, indicating a more heterogeneous distribution of population density in the former as compared to the latter. 

\subsection{Model}

To characterize a city's vulnerability to disease spread, we calculate the epidemic threshold by generalizing the formalism introduced in~\cite{Gomez-Gardenes2018}. In what is to follow, we incorporate a Susceptible (S), infectious (I) or recovered (R) dynamics, however, we note that the formalism is easily generalizable to more elaborate compartmental schemes. An infectious individual transmits the pathogen to healthy counterparts via direct interaction at a rate $\lambda$. In turn, infectious individuals enter the compartment $R$ at a rate $\mu$, which typically encodes the inverse of the expected contagious period. The mixing among healthy and infectious individuals is governed by the spatial distribution of the population and their mobility patterns. The different zip-codes of the city are represented as patches $i$, which are initially populated by $n_i$ residents.  The activity of the residents is considered on a daily basis and split into three stages: Movement, Interaction and Return. Residents decide to either move to a different patch with probability $p$, or remain with probability $1-p$. If the former, the choice of location is proportional to the elements of the origin destination matrix ${\bf T}$. After all movements (or lack thereof) have been completed, interactions occur within patches according to a mean-field assumption where every individual makes the same number of contacts proportional to the population density via a function $f_{i}$; the final step involves return to the place of residence. The same process is then repeated for the next time-step (day). 

Given $N$ patches in a city,  the dynamics is completely specified by $2\times N$ coupled discrete equations governing the temporal evolution of the fraction of infected and recovered individuals residing in each patch. Namely, the fraction of infected, $\rho^{I}_i(t+1)$, and recovered individuals, $\rho^{R}_{i}(t+1)$, associated to patch $i$ at time $t+1$, reads:
\begin{eqnarray}
\rho^I_i(t+1)&=& (1-\mu) \rho^I_i (t) + (1-\rho^I_i(t)-\rho^R_i(t))\Pi_i (t)\ ,\label{eq:rho} \nonumber\\
\rho_i^R(t+1)&=& \rho^R_i(t)+ \mu\rho^I_i (t),\ 
\label{eq:rhor}
\end{eqnarray}
where $\Pi_i (t)$  represents the probability of a susceptible individual resident in $i$ to contract the disease at time $t$. Assuming that the process reaches a steady-state and that the size of the outbreak is small in comparison to the overall population, after a sequence of manipulations (Supplementary Section S3, Eqns. S4-S15), it can be shown that the epidemic threshold  is of the form, $\lambda_c (p) = \mu /\Lambda_{\text {max}} \left({\bf M}\right)$, where $\Lambda_{\text{max}}$ is the spectral radius of a mixing matrix ${\bf M}$, taking into account the mobility flows, the degree of mobility $p$, the effective population in a given patch, and the number of contacts as a function of population density in that patch. If $p=0$ then the threshold would correspond to a static population that never moves, whereas if $p=1$ then this accounts for a fully active population.
Based on this observation one can define a normalized epidemic threshold $\tilde{\lambda_c} =\lambda_c(p=1)/\lambda_c(p=0)$ to focus only on mobility effects while removing the influence of the population density.

\subsection{Connecting the epidemic threshold to hotspot flows}

Given the quantitative description of human flows composing the mobility backbone of each city, we now focus on determining the effect that their morphology has on the epidemic threshold. Given the extensive list of considered cities, and the attendant variation in population density, in order for a fair characterization, it is important to remove this component, and therefore the relevant parameter here is the normalized threshold $\tilde{\lambda}_c$.
\begin{figure}[t!]
    \centering
   \includegraphics[width=0.86\columnwidth]{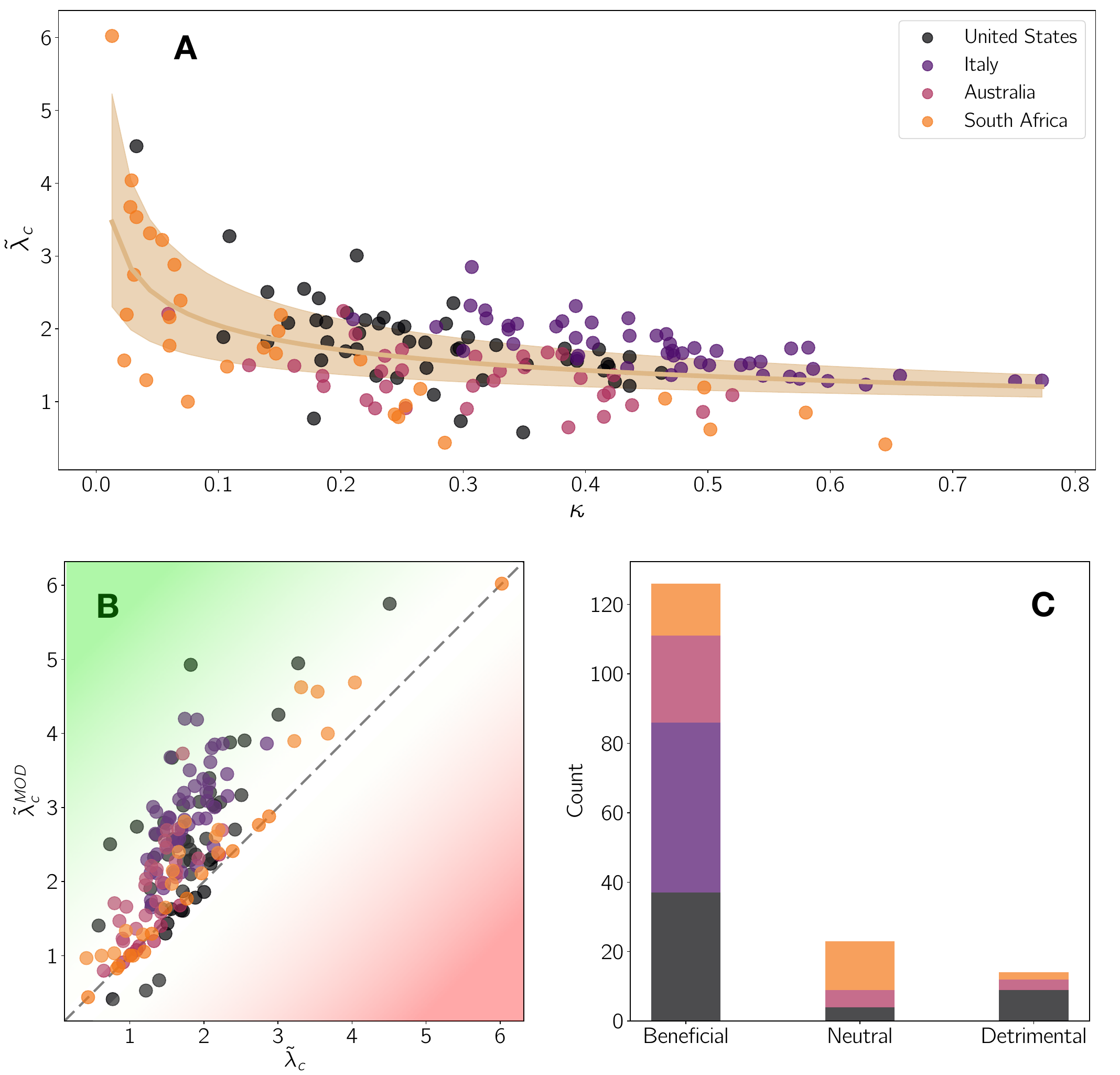}
    \caption{{\bf Connecting morphology to vulnerability} {\bf A} Normalized epidemic threshold $\tilde{\lambda}_c$ versus population density hotspot concentration $\kappa$ for cities in each of the chosen 4 countries (color code). The solid line shows the fit to a power-law function given by $\tilde{\lambda}_c = A\kappa^{\beta}$, with $A=1.15 \pm 0.06$ and $\beta= -0.25 \pm 0.03$. The shadowed region covers the 95\% confidence interval. {\bf B} Normalized threshold after removing all the flows connecting hotspots $\tilde{\lambda}_c^{MOD}$ as a function of the normalized epidemic threshold without intervention $\tilde{\lambda}_c$. The green (red) areas contains those cities for which removing flows among hotspots is beneficial (detrimental). {\bf C} Division of the cities according to the result: beneficial, neutral and detrimental as described in the text.}
    \label{fig:hotspot_flow}
\end{figure}
In Fig.~\ref{fig:hotspot_flow}{\bf A} we plot $\tilde \lambda_c$ as a function of $\kappa$ for all the cities in our dataset, represented as filled circles colored according to their respective countries. We find a universal trend, whereby all cities fall into a curve marking an inverse relationship between the vulnerability of the cities and the extent to which flows are concentrated between hotspots, i.e $\tilde \lambda_c \sim \kappa^{\beta}$ with $\beta \approx -0.25$. In particular, cities where there is a more egalitarian distribution of flows (lower $\kappa$), the epidemic threshold is higher for a mobile population as compared to a static population, indicating that movement between regions lowers the risk of an epidemic outbreak. Conversely, in those cities where the population moves primarily between hotspots, there is little-to-no difference in risk in terms of whether residents stay in their patches, or whether they move to different ones. In Fig. S1, we show the equivalent of Fig.~\ref{fig:hotspot_flow}{\bf A} but now split by country finding the same trend attesting to the robustness of the inverse dependence. (The fits to $\beta$ for each country are shown in Tab. S1). As a comparison to other population-related measures, in Fig. S2, we plot the Spearman correlation matrix for $\kappa$, the average population density, the Lloyd mean crowding~\cite{Scarpino2020} and $\tilde{\lambda_c}$ for the 50 American cities. The figure suggests that the concentration of flows between areas of high population density, $\kappa$, is a much better predictor of the vulnerability of cities than the other two measures. 

To further characterize the relation existing between cities' vulnerability and the concentration of mobility between density hotspots, we next analyze the impact of reshuffling the flows at a local scale for each of the analyzed cities. In particular, we preserve the total amount of flows, by removing all links connecting hotspots (thus setting $\kappa = 0$) and redistributing them evenly across non-hotspot locations. We then recompute the resulting normalize threshold $\tilde{\lambda}_c^{MOD}$ and in Fig.~\ref{fig:hotspot_flow}{\bf B} plot it as a function of the threshold corresponding to the unperturbed network $\tilde{\lambda}_c$. As the figure indicates for the vast majority of cities (irrespective of their original value of $\kappa$), the effect of switching off the flows between hotspots leads to an increase in the epidemic threshold, in turn lowering their vulnerability to epidemic spread. In particular there appears to be three categories of cities: {\it  beneficial}, those where the threshold is increased greater than $10\%$; {\it neutral}, those where the threshold is increased by less than $10\%$; and finally {\it detrimental}, those where the threshold is instead lowered. The number of cities belonging to each category is plotted as a bar-chart in Fig.~\ref{fig:hotspot_flow}{\bf C}, indicating that around $75\%$ of cities experience a lowering of their vulnerability, $~15\%$ remain neutral, and the remaining $10\%$ experience an increased susceptibility (interestingly this category is dominated by American cities). For the small number of cities, where we find this counter-intuitive effect of lowered threshold, it is likely that there are other more complex features at play not considered in this analysis. 

Nevertheless, these results provide overwhelming evidence that, in most cases, the concentration of human mobility between densely populated areas is a feature that enhances disease spreading and makes such cities vulnerable to epidemics. Moreover, the beneficial effect caused by the reorientation of intra-hotspot flows towards less densely populated areas seems to be rooted in a homogenization of the distribution of the underlying density, which, as mentioned in~\cite{Soriano-Panos2018,Soriano-Panos2020}, enforces infected individuals to stay away from the contagion focus, thus reducing their infection power. In turn, this homogenizing flow structure appears naturally in cities with low $\kappa$ and, as suggested by the empirical trends in Fig. ~\ref{fig:hotspot_flow} {\bf A}, characterizes the most resilient cities. Therefore, a lower $\kappa$ translates into a greater mix of populations, between high and low population density centers, where they can actually take advantage of mobility between city sub-regions to prevent outbreaks.

\subsection{Application to real pandemic settings}
The formalism proposed here can readily be applied to assess the exposure of cities to actual outbreaks. To illustrate this, we next focus on the  spread of SARS-COV-2 in the 50 most populated Core-Based-Statistical-Areas (CBSA) in the United States, chosen due to the appropriate spatial resolution in terms of infection data. Note that, although, thus far, we have focused on the SIR model, the following will illustrate the generality of the results, in the context of the spread of SARS-CoV-2, that has been recently analyzed with more elaborate compartmental models~\cite{Colizza2020,prem2020effect,Scarpino2020,Arenas2020,gatto2020spread}.


As a proxy for a city's vulnerability to epidemic spread, we make use of the number of confirmed infected cases at the county-level collected from the New York Times \footnote{\url{https://github.com/nytimes/covid-19-data}} and USAFacts \footnote{\url{https://usafacts.org/visualizations/coronavirus-covid-19-spread-map/}}. Given the inherent noise due to reporting artifacts, and assuming an exponential growth, $I_k(t) \sim \exp(b_k t)$ during early onset, we apply a smoothing procedure to extract the growth-rate $b_k$ of the number of infected cases, and use that as proxy for a city $k$'s susceptibility to disease spread. To remove any effects due to non-pharmaceutical interventions and behavioral changes in the population, we focus on the period before mitigation measures. The full details of the procedure are shown in Supplementary Section S5, and the temporal infection plots for each city along with the fits are shown in Fig. S3.

\begin{figure}[t!]
     \centering
     \includegraphics[width=1\columnwidth]{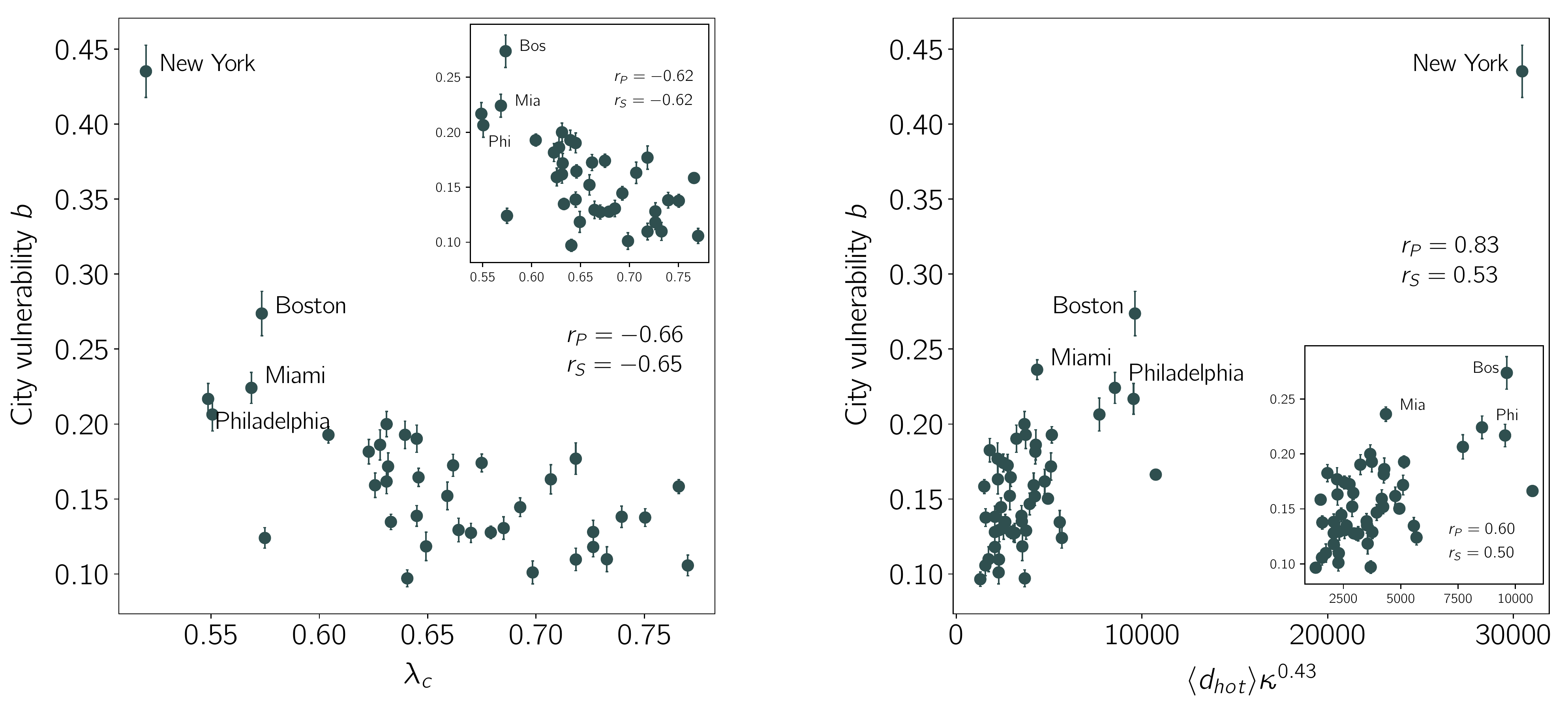}
     \caption{{\bf Validation of the model in US cities} {\bf A} Epidemic growth rate $b$ as a function of the epidemic threshold $\lambda_c (p)$ (Eq. S16). In inset we show the same plot after removing NYC. {\bf B} Epidemic growth rate $b$ as a function of the average population density within hotspots $\langle d_{hot}\rangle$ and the hotspot concentration $\kappa$. The exponent $0.43$ is introduced to reflect the dependence obtained when fitting the  normalized epidemic threshold to $\kappa$ for the United States (Table S2). In both panels, $r_P$ and $r_S$ denote the Pearson and Spearman rank correlation coefficients among the represented quantities.}
     \label{fig:infection_indicators}
\end{figure}

In order to properly connect with the growth rate, we need to reintroduce the effect of the population distribution and, therefore, the relevant variable is the unnormalized threshold $\lambda_c (p=1)$. Note that, by choosing $p=1$, we force all the inhabitants within a city to follow the flow matrix $\bf T$, but not all of them leave their residential area due to existence of self-loops in this matrix. Indeed, according to the avalaible data for the cities analyzed here, around $36\%$ of the population remains on average inside their residential administrative unit.

In Fig. \ref{fig:infection_indicators}{\bf A} we plot the empirically extracted growth rate $b$ as a function of the epidemic threshold, $\lambda_c$, finding once again an inverse trend, confirming the role of the threshold as a proxy for vulnerability. Those cities which experienced a faster epidemic growth during the early onset of the pandemic indeed had a lower threshold according to our formalism.



Next we connect, the localization of flows to hotspots to the empirical vulnerability in each of the cities. Since $\kappa$ only takes into account flows between hotspots but does not account for their population density, the variable that captures the effective interaction between the residents in hotspots areas should be a combination of both factors. Specifically, we have chosen $\langle d_{hot} \rangle \kappa^{0.43}$, where the first term corresponds to the average population density within hotspots and the second term reflects the scaling obtained for the normalized epidemic threshold in the individual case of CBSA from the United States (see Supplementary Section 4 for further details). In Fig. \ref{fig:infection_indicators}{\bf B} we plot $b$ as a function of this quantity finding a clear monotonically increasing trend. Thus taken together, the results of Fig. \ref{fig:infection_indicators} indicate that the empirical trends mirror our theoretical formalism, whereby cities that experience strong early growth of the epidemic have a lower threshold, a phenomenon for which one of the main causative mechanisms is that the movement in cities occurs primarily between hotspots.



\subsection{Potential mitigation measures}

The observations thus far, immediately suggest the possibility of effective mitigation measures that may shore up the robustness of vulnerable cities to the onset of epidemic spread. Given the lack of therapeutics or vaccines for SARS-COV-2, the prevailing strategy adopted globally has been to resort to non-pharmaceutical interventions of which a key ingredient has been aggressive lock-downs. While ostensibly being very effective in mitigating an active epidemic, significant disruption to the socio-economic fabric is one of the unfortunate consequences \cite{eurostat2020,bea2020}. Having demonstrated the key role played by the interactions between population density hotspots, we next investigate some targeted interventions, or even preemptions, that are milder than completely restricting mobility city-wide and assess their efficacy in reducing vulnerability. The strategy we pursue is to modify flows between different types of locations in the city without the need to isolate individuals at home. The different schemes are illustrated in the upper panel of Fig. 4. In the first intervention (Intervention I), an asymmetrical strategy involves restricting flows from non-hotspot (heretofore referred to as suburbs) towards hotspot areas and converting them to self-loops, while keeping all other flows the same. Intervention II corresponds to the reverse situation where flows from hotspots to suburbs are converted to self-loops. Finally, in Intervention III, only movement between hotspots is restricted.

\begin{figure}[t!]
    \centering
    \includegraphics[width=0.92\columnwidth]{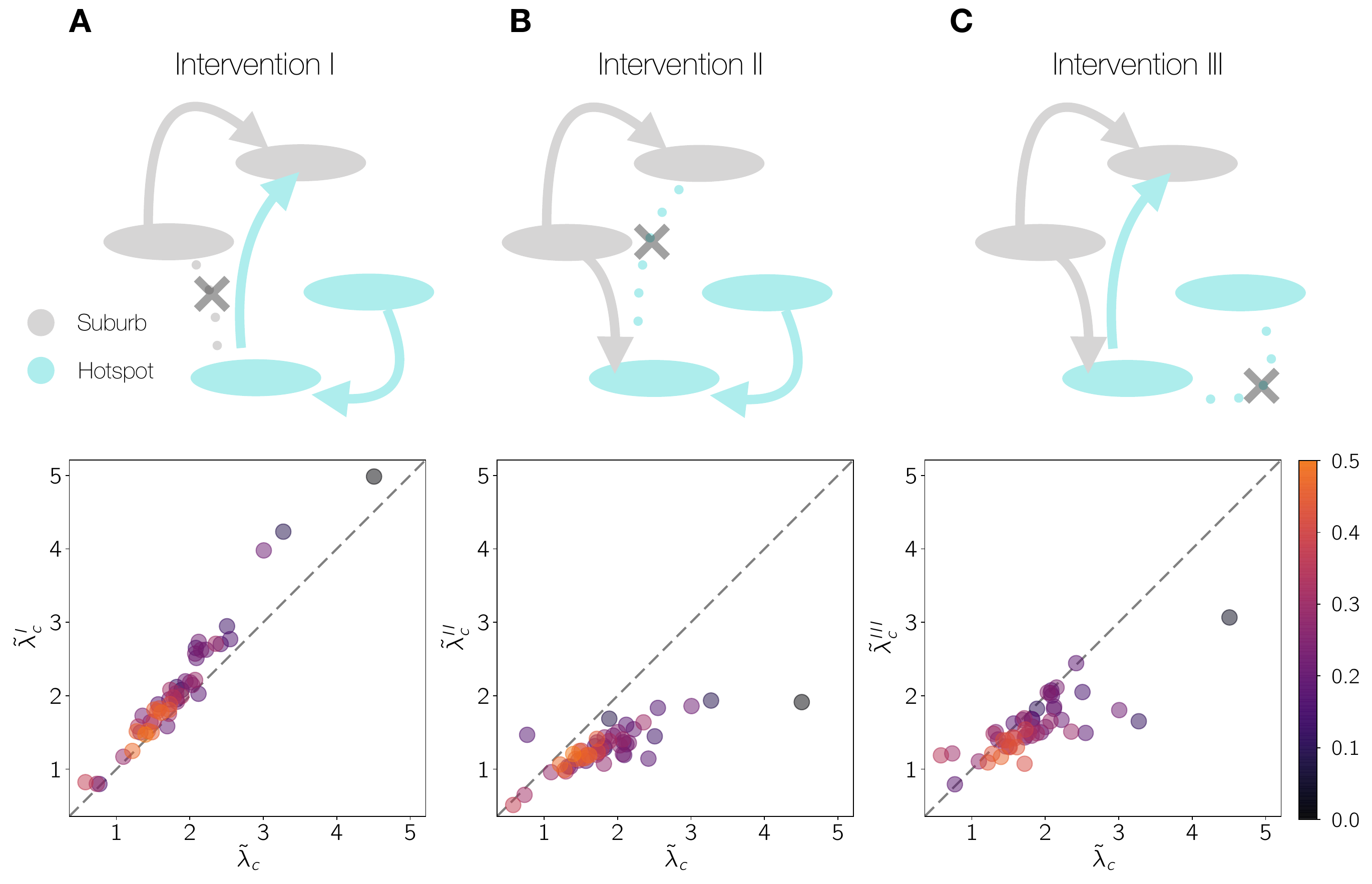}
    \caption{{\bf Impact of mobility interventions on the epidemic threshold within cities in the United States}. The impact is quantified by representing the normalized threshold for each intervention, $\tilde{\lambda}_c^{I}$, $\tilde{\lambda}_c^{II}$, $\tilde{\lambda}_c^{III}$ (in panels {\bf A,B,C} respectively) as a function of the normalized threshold $\tilde{\lambda}_c$. In all panels, the dashed line denotes the boundary separating the cities for which the intervention is beneficial (above) or detrimental (below). The color of the points represent the hotspot concentration $\kappa$ for each CBSA.}
    \label{fig:scenarios}
\end{figure}


For each of these scenarios we recompute the normalized epidemic threshold $\tilde{\lambda}_c^{m}$ ($m\in \text{\{I,II,III\}}$) and plot it against the original threshold, $\tilde{\lambda_c}$, in the bottom panels of Fig. 4. In panel {\bf A}, we find an increase of the threshold across the board indicating that Intervention I is a rather effective strategy. Given that a potential epidemic has a high probability of being seeded and correspondingly spreading extensively in high population density areas, preventing the residents in suburbs from visiting these locations protects them from being exposed to the disease. Conversely in panel {\bf B} we see that restricting residents in hotspots to travel to suburbs has the opposite effect in further \emph{decreasing} the threshold. This counterproductive effect emerges due to a phenomenon discussed in~\cite{Soriano-Panos2018}. When there is an asymmetry between the population density in cities, mobility from hotspots to suburbs leads to an increase in the threshold due to the dilution of the effective population in the hotspots thus reducing the number of contacts, as well as diverting potentially infectious individuals to to lower density regions where their impact is mitigated. Removing this route makes the situation significantly worse. Finally, panel {\bf C} reveals that limiting the mobility between hotspots has a mostly neutral effect, although the trends are noisier given that its effectiveness is closely related to the underlying population density distribution inside hotspots. The results suggest that, among the scenarios outlined, the most beneficial policy is to restrict residents in suburbs from visiting hotspots, while at the same time allowing residents in hotspots to continue with their regular mobility behavior. Note, that several other combinations are possible, for instance a combination of interventions I \& III which is likely to have even more of a beneficial effect.

\section{Discussion}

Similar to how a virus enters the human body and replicates from cell to cell, the spread of pathogens in susceptible populations is influenced by the interaction between its hosts. Thus, it is the social interactions, mediated by behavioral and mixing patterns, that shapes the spread of disease in human populations. Among these aspects, human mobility is a key factor underlying the unfolding patterns of epidemic outbreaks. Understanding how human mobility shapes the spatiotemporal unfolding of contagious diseases is essential for the design of efficient containment policies to ameliorate their impact. In this paper, we investigate the interplay between the population density and the spatial distribution of flows in urban areas, and its impact on determining exposure to epidemic outbreaks. We report a clear trend worldwide: the existence of a high volume of individuals commuting among the population density centers of a given city makes it more vulnerable to the spread of epidemics. The extent of a city's vulnerability as determined from our formalism, allows us to shed light on a real epidemic scenario: the spread of SARS-COV-2 across United States. In particular, the epidemic threshold determined by the population density profiles and urban mobility patterns provides one of the potential causative mechanisms behind the different levels of infection observed across cities in the United States. 

As an application to potential epidemic waves, our indicator allows for identifying those cities that are likely to become epidemic centers once the first imported cases arrive there. This is of importance, given that it can guide authorities to identify places where timely containment policies can be locally implemented to avoid large outbreaks caused by massive community transmission. It is precisely the lack of anticipation to the SARS-COV-2 pandemic, that has led countries to enforce aggressive containment measures aimed at ameliorating the impact of the disease. The predominant strategy has been the implementation of lockdown policies, forcing a large fraction of the population to stay isolated at home, thus reducing considerably their number of interactions. While there is consensus on the effectiveness of these interventions to mitigate an ongoing outbreak, the collateral socio-economic damage caused by lockdowns requires a change of direction towards less-aggressive containment measures. 

In the case of SARS-COV-2, the strict individual isolation characterizing the first interventions have given rise to more relaxed lockdown scenarios combined with efficient Test-Track and Isolate policies \cite{hellewell2020feasibility,ferretti2020quantifying,salathe2020covid}. Along these lines, we present different realistic scenarios based on modifying mobility habits to actively avoid the emergence of large areas of contagion. Our analysis suggest that a potentially effective policy involves an asymmetric closure of neuralgic centers of the cities, restricting movement to population density hotspots from residents of other areas, while allowing those living in hotspots to commute, in order to dilute the number of contacts in the most vulnerable areas. Teleworking and effective distribution of key services in a city are practical manifestations of such interventions.

An advantage of our formalism is its relative simplicity, paving the way to extend the results to more general scenarios. For example, these results are built upon the assumption that population density centers are much more vulnerable to contagious diseases than scarcely dense areas. While this is a logical assumption, the beneficial effect of mobility from hotspots and suburbs could be reversed for diseases with large reproduction number $R_0$. In this scenario, suburbs would also have the potential to develop large outbreaks so the existence of infection routes across the city would lead to an acceleration of the propagation of the epidemic front.  Finally, another obvious extension is to consider movement at different geographical scales~\cite{Wattsmultiscale,balcan2009multiscale}. Accounting for movement between cities, for instance, could provide valuable information to coordinate joint efforts among different regions to modify both inter- and intra-urban flows in service of reducing the impact of a pandemic. Needless to say, pandemics are complex processes involving a multitude of spatial and socioeconomic factors. The results presented here may provide one of the building blocks for policy planners in devising effective preventive and mitigation measures for future crises.

\noindent {\bf Limitations}

These results should be interpreted in light of several important limitations. First, the Google mobility data is limited to smartphone users who have opted in to Google’s Location History feature, which is off by default. These data may not be representative of the population as whole, and furthermore their representativeness may vary by location. Importantly, these limited data are only viewed through the lens of differential privacy algorithms, specifically designed to protect user anonymity and obscure fine detail. Moreover, comparisons across rather than within locations are only descriptive since these regions can differ in substantial ways.
\bibliographystyle{naturemag}
\bibliography{references}
\newpage
{\bf	\Large{Supporting Information}}
\renewcommand{\thefigure}{S\arabic{figure}}
\renewcommand{\thesection}{S\arabic{section}}
\renewcommand{\thetable}{S\arabic{table}}
\renewcommand{\theequation}{S\arabic{equation}}
\renewcommand{\thepage}{S-\arabic{page}}

\setcounter{section}{0}
\setcounter{equation}{0}
\setcounter{figure}{0}
\setcounter{page}{1}
\section{Hotspot classification}
Hotspots are identified by setting a threshold on the population densities of cells within a city. The threshold for hotspots is assigned by applying a non-parametric method, the LouBar method~\cite{Louail2014,Bassolas2019}, based on the derivative of the Lorenz curve. The Lorenz curve is the sorted cumulative distribution of population densities and is obtained by plotting, in ascending order, the normalized cumulative number of nodes vs. the normalized cumulative population density. The threshold is then obtained by taking the derivative of the Lorenz curve at (1, 1) and extrapolating it to the point at which it intersects the x-axis. 
We classify hotspots in cities according to this LouBar method applied to population densities, in agreement with the reliance of the model on effective densities. A cell $i$ is considered a hotspot of city $k$, $H_k$ if it satisfies:
\begin{equation}
    d_{i, k} > d^{\text{Lou}}_k
\end{equation}
where $d_i$ is the population density of cell $i$ in city $k$ and $d^{\text{Lou}}_k$ is the threshold determined by performing the LouBar method on the population densities of all cells in city $k$. This allows us to place emphasis on zones within cities that encourage the most relative interaction, as opposed to sharp biasing due to population magnitude. We then examine the hotspots flow concentration in each city $k$, $\kappa_k$, defined as fraction of total flows in the city system that exist between these population density hotspots of city. Therefore, $\kappa_k$ is given by:
\begin{equation}
    \kappa_k = \frac{\sum\limits_{i, j \epsilon H_k}{T_{ij}}}{\sum\limits_{i,j}T_{ij}}\ ,
\end{equation}
where $T_{ij}$ denotes the flow of individuals going from patch $i$ to patch $j$ according to the mobility data. 

\section{Data}

 \begin{filecontents*}{Citiestable.csv}
Citynew,Country,Resolution,Hotspotflow
Bunbury,AUS,SA2 within SA4,0.059
Capital Region,AUS,SA2 within SA4,0.125
Sydney ,AUS,SA2 within SA4,0.162
Darwin,AUS,SA2 within SA4,0.185
Cairns,AUS,SA2 within SA4,0.186
Richmond ,AUS,SA2 within SA4,0.202
Gold Coast,AUS,SA2 within SA4,0.212
South Australia ,AUS,SA2 within SA4,0.221
Hume,AUS,SA2 within SA4,0.228
Moreton Bay ,AUS,SA2 within SA4,0.233
Central Queensland,AUS,SA2 within SA4,0.236
Wide Bay,AUS,SA2 within SA4,0.237
Melbourne ,AUS,SA2 within SA4,0.250
Townsville,AUS,SA2 within SA4,0.250
Australian Capital Territory,AUS,SA2 within SA4,0.253
Launceston and North East,AUS,SA2 within SA4,0.303
Sunshine Coast,AUS,SA2 within SA4,0.308
Central Coast,AUS,SA2 within SA4,0.310
Illawarra,AUS,SA2 within SA4,0.325
Perth ,AUS,SA2 within SA4,0.330
Logan ,AUS,SA2 within SA4,0.349
Brisbane ,AUS,SA2 within SA4,0.350
Hunter Valley exc Newcastle,AUS,SA2 within SA4,0.369
Newcastle and Lake Macquarie,AUS,SA2 within SA4,0.381
Western Australia ,AUS,SA2 within SA4,0.386
Ipswich,AUS,SA2 within SA4,0.396
West and North West,AUS,SA2 within SA4,0.415
Mackay ,AUS,SA2 within SA4,0.415
Latrobe ,AUS,SA2 within SA4,0.419
North West,AUS,SA2 within SA4,0.423
Hobart,AUS,SA2 within SA4,0.438
Central West,AUS,SA2 within SA4,0.496
Adelaide ,AUS,SA2 within SA4,0.520
Bolzano,ITA,S2 cells within communes,0.210
Sassari,ITA,S2 cells within communes,0.278
Catania,ITA,S2 cells within communes,0.300
Livorno,ITA,S2 cells within communes,0.306
Taranto,ITA,S2 cells within communes,0.307
Foggia,ITA,S2 cells within communes,0.318
Perugia,ITA,S2 cells within communes,0.319
Bari,ITA,S2 cells within communes,0.337
Giugliano In Campania,ITA,S2 cells within communes,0.337
Terni,ITA,S2 cells within communes,0.341
Piacenza,ITA,S2 cells within communes,0.344
Genova,ITA,S2 cells within communes,0.376
Ravenna,ITA,S2 cells within communes,0.381
Ferrara,ITA,S2 cells within communes,0.392
Ancona,ITA,S2 cells within communes,0.392
Lecce,ITA,S2 cells within communes,0.393
Cesena,ITA,S2 cells within communes,0.394
Pesaro,ITA,S2 cells within communes,0.405
Parma,ITA,S2 cells within communes,0.406
Pescara,ITA,S2 cells within communes,0.434
Roma,ITA,S2 cells within communes,0.435
Venezia,ITA,S2 cells within communes,0.436
Modena,ITA,S2 cells within communes,0.458
Trieste,ITA,S2 cells within communes,0.466
Brescia,ITA,S2 cells within communes,0.467
Trento,ITA,S2 cells within communes,0.469
Siracusa,ITA,S2 cells within communes,0.470
Napoli,ITA,S2 cells within communes,0.471
Palermo,ITA,S2 cells within communes,0.472
Milano,ITA,S2 cells within communes,0.478
Rimini,ITA,S2 cells within communes,0.479
Prato,ITA,S2 cells within communes,0.489
Verona,ITA,S2 cells within communes,0.494
Salerno,ITA,S2 cells within communes,0.501
Torino,ITA,S2 cells within communes,0.507
Arezzo,ITA,S2 cells within communes,0.527
Reggio Di Calabria,ITA,S2 cells within communes,0.534
Latina,ITA,S2 cells within communes,0.543
Udine,ITA,S2 cells within communes,0.545
Bologna,ITA,S2 cells within communes,0.567
Vicenza,ITA,S2 cells within communes,0.568
Andria,ITA,S2 cells within communes,0.575
Novara,ITA,S2 cells within communes,0.582
Padova,ITA,S2 cells within communes,0.586
Bergamo,ITA,S2 cells within communes,0.598
Firenze,ITA,S2 cells within communes,0.629
Messina,ITA,S2 cells within communes,0.657
Cagliari,ITA,S2 cells within communes,0.751
Monza,ITA,S2 cells within communes,0.773
Virginia Beach,USA,Zip codes within CBSA,0.033
Washington,USA,Zip codes within CBSA,0.104
Miami,USA,Zip codes within CBSA,0.109
Columbus,USA,Zip codes within CBSA,0.140
Seattle,USA,Zip codes within CBSA,0.140
Atlanta,USA,Zip codes within CBSA,0.157
Houston,USA,Zip codes within CBSA,0.170
Pittsburgh,USA,Zip codes within CBSA,0.178
Richmond,USA,Zip codes within CBSA,0.180
Raleigh,USA,Zip codes within CBSA,0.182
San Francisco,USA,Zip codes within CBSA,0.184
Nashville,USA,Zip codes within CBSA,0.188
Los Angeles,USA,Zip codes within CBSA,0.189
Salt Lake City,USA,Zip codes within CBSA,0.204
Cincinnati,USA,Zip codes within CBSA,0.205
New York,USA,Zip codes within CBSA,0.213
Austin,USA,Zip codes within CBSA,0.213
Minneapolis,USA,Zip codes within CBSA,0.215
Birmingham,USA,Zip codes within CBSA,0.220
Boston,USA,Zip codes within CBSA,0.229
Providence,USA,Zip codes within CBSA,0.231
Charlotte,USA,Zip codes within CBSA,0.235
Louisville/Jefferson County,USA,Zip codes within CBSA,0.246
Dallas,USA,Zip codes within CBSA,0.247
Orlando,USA,Zip codes within CBSA,0.252
St. Louis,USA,Zip codes within CBSA,0.256
Riverside,USA,Zip codes within CBSA,0.269
Buffalo,USA,Zip codes within CBSA,0.270
Chicago,USA,Zip codes within CBSA,0.276
Kansas City,USA,Zip codes within CBSA,0.286
Baltimore,USA,Zip codes within CBSA,0.292
San Jose,USA,Zip codes within CBSA,0.295
Denver,USA,Zip codes within CBSA,0.297
Philadelphia,USA,Zip codes within CBSA,0.298
Cleveland,USA,Zip codes within CBSA,0.304
Hartford,USA,Zip codes within CBSA,0.316
San Antonio,USA,Zip codes within CBSA,0.327
Portland,USA,Zip codes within CBSA,0.349
San Diego,USA,Zip codes within CBSA,0.352
Phoenix,USA,Zip codes within CBSA,0.383
Milwaukee,USA,Zip codes within CBSA,0.385
Memphis,USA,Zip codes within CBSA,0.393
Indianapolis,USA,Zip codes within CBSA,0.411
Las Vegas,USA,Zip codes within CBSA,0.415
Oklahoma City,USA,Zip codes within CBSA,0.418
Sacramento,USA,Zip codes within CBSA,0.419
New Orleans,USA,Zip codes within CBSA,0.424
Jacksonville,USA,Zip codes within CBSA,0.436
Tampa,USA,Zip codes within CBSA,0.436
Detroit,USA,Zip codes within CBSA,0.462
Govan Mbeki,ZAF,Wards within municipalities,0.013
Hibiscus Coast,ZAF,Wards within municipalities,0.023
City of Matlosana,ZAF,Wards within municipalities,0.025
Nelson Mandela Bay,ZAF,Wards within municipalities,0.028
City of Cape Town,ZAF,Wards within municipalities,0.029
Emalahleni,ZAF,Wards within municipalities,0.031
City of Tshwane,ZAF,Wards within municipalities,0.033
Local Municipality of Madibeng,ZAF,Wards within municipalities,0.041
Ekurhuleni,ZAF,Wards within municipalities,0.044
City of Johannesburg,ZAF,Wards within municipalities,0.054
Matjhabeng,ZAF,Wards within municipalities,0.060
Nkomazi,ZAF,Wards within municipalities,0.060
Greater Letaba,ZAF,Wards within municipalities,0.064
Mangaung,ZAF,Wards within municipalities,0.069
Makhado,ZAF,Wards within municipalities,0.075
Emfuleni,ZAF,Wards within municipalities,0.107
Greater Tzaneen,ZAF,Wards within municipalities,0.137
eThekwini,ZAF,Wards within municipalities,0.147
Mogalakwena,ZAF,Wards within municipalities,0.149
Lepele,ZAF,Wards within municipalities,0.151
Buffalo City,ZAF,Wards within municipalities,0.216
Polokwane,ZAF,Wards within municipalities,0.244
The Msunduzi,ZAF,Wards within municipalities,0.247
Rustenburg,ZAF,Wards within municipalities,0.253
Makhuduthamaga,ZAF,Wards within municipalities,0.265
Thembisile,ZAF,Wards within municipalities,0.285
Chief Albert Luthuli,ZAF,Wards within municipalities,0.465
Dr JS Moroka,ZAF,Wards within municipalities,0.497
Maluti a Phofung,ZAF,Wards within municipalities,0.502
Bushbuckridge,ZAF,Wards within municipalities,0.580
Thulamela,ZAF,Wards within municipalities,0.645
\end{filecontents*}
\csvreader[
   longtable=lclc,
   table head=\caption{Hotspot flow concentration $\kappa$ for the different cities analyzed in the manuscript. The resolution column contain the two geographical divisions used inside each country to construct the metapopulations. For example, ``Zip codes within CBSA" in the case of the USA implies that each city (metapopulation) corresponds to a CBSA and the entities composing each city (patches) correspond to zip codes. \label{tab:sometab}}\\
     \toprule\bfseries City &\bfseries Country &\bfseries Resolution & \bfseries $\kappa$ \\ \midrule\endfirsthead
    \toprule\bfseries City &\bfseries Country &\bfseries Resolution & \bfseries $\kappa$ \\ \midrule\endhead
   \bottomrule\endfoot,
   late after line=\\,
]{Citiestable.csv}{1=\City, 2=\Country, 3=\Resolution, 4=\Hotspotflow}
{\City&\Country&\Resolution&\Hotspotflow}


\subsection{Mobility Data}
The Google COVID-19 Aggregated Mobility Research Dataset contains anonymized mobility flows aggregated over users who have turned on the Location History setting, which is off by default. This is similar to the data used to show how busy certain types of places are in Google Maps --- helping identify when a local business tends to be the most crowded. The dataset aggregates flows of people from region to region.

To produce this dataset, machine learning is applied to logs data to automatically segment it into semantic trips \cite{Bassolas2019}. To provide strong privacy guarantees, all trips were anonymized and aggregated using a differentially private mechanism \cite{googleresearch} to aggregate flows over time (see \url{https://policies.google.com/technologies/anonymization}). This research is done on the resulting heavily aggregated and differentially private data. No individual user data was ever manually inspected, only heavily aggregated flows of large populations were handled.

All anonymized trips are processed in aggregate to extract their origin and destination location and time. For example, if  users traveled from location $a$ to location $b$ within time interval $t$, the corresponding cell $(a,b,t)$  in the tensor would be $n \pm \textrm{err}$, where err is Laplacian noise. The automated Laplace mechanism adds random noise drawn from a zero mean Laplace distribution and yields $(\epsilon,\delta$)--differential privacy guarantee of $\epsilon = 0.66$ and $\delta =2.1\times 10^{-29}$ per metric. Specifically, for each week $W$ and each location pair $(a,b)$, we compute the number of unique users who took a trip from location $a$ to location $b$ during week $W$. To each of these metrics, we add Laplace noise from a zero-mean distribution of scale $1/0.66$. We then remove all metrics for which the noisy number of users is lower than 100, following the process described in \footnote{https://research.google/pubs/pub48778/}, and publish the rest. This yields that each metric we publish satisfies $(\epsilon,\delta$)--differential privacy with values defined above. The parameter $\epsilon$ controls the noise intensity in terms of its variance, while $\delta$ represents the deviation from pure $\epsilon$--privacy. The closer they are to zero, the stronger the privacy guarantees. 

We use data collected weekly from November 3rd 2019 to February 29th 2020, ensuring that we capture standard movement behavior, uninfluenced by pandemic conditions. Depending on the availability of population data per country, the mobility flows between S2 cells are aggregated to patch size of comparable resolution between countries:

\begin{itemize}
    \item United States (USA): 50 Urban areas with zip code patches~\cite{us_pop}.
    \item Italy (ITA): 49 Communes with S2 cell patches~\cite{fb_pop}.
    \item South Africa (ZAF): 31 Municipalities with ward patches~\cite{zaf_gov}.
    \item Australia (AUS): 33 Statistical Areas Level 4, with Level 2 patches~\cite{aus_gov}.
\end{itemize}

To aggregate the S2 cells to the corresponding alternate patch types, the centroid points of the S2 cells were spatially joined with GIS boundaries of the alternate resolution. For example, in the US, flows from a zip code X to a zip code Y are determined by summing all of the flows that start in S2 cells whose centroids lie within X and end in S2 cells whose centroids lie within Y. 

\subsection{Population density data}
Population density data is aggregated to patch-size in a similar manner when necessary. In the case of US zip codes, Australian statistical areas, and South African wards, population information is collected and available at those respective resolutions. For patches in Italy, population is determined by aggregating the populations of the 30 meter tiles collected by Facebook \cite{fb_pop} to S2 cells. This is done by spatially merging the centroids of the tiles to the patches, similar to the S2-to-patch mobility flow aggregation.

\section{Epidemic model}
\subsection{Dynamical equations}
The model used for estimating the vulnerability of a given city is a generalized version of the formalism included in \cite{Gomez-Gardenes2018}. From an epidemiological point of view, the model incorporates a SIR dynamics where individuals can be susceptible of contracting the disease (S), infectious (I) or recovered (R). An infectious individual transmits the pathogen to healthy counterparts via direct interaction at a rate $\lambda$. In turn, infectious individuals enter the compartment $R$ at a rate $\mu$, which typically encodes the inverse of the expected contagious period. The mixing among healthy and infectious individuals is governed by the spatial distribution of the population and their mobility patterns, which are accommodated in the formalism by using metapopulations. A metapopulation is a complex network composed by a set of $N$ patches which represent places gathering agents. Those individuals can move across the metapopulation, being these movements determined by the mobility patterns usually encoded in origin-destination matrices. 

In our model, each individual has an associated node which is identified as her residence. Therefore, each patch $i$ is initially populated by $n_i$ agents. We split each day into three stages: Movement, Interaction and Return. First, agents decide whether or not moving with a probability $p$. If moving, they choose their destination according to the origin destination matrix ${\bf R}$, being the elements $R_{ij}$ the probability of moving from patch $i$ to patch $j$. We construct this matrix from mobility data, encoded in {\bf T}, as
\begin{equation}
    R_{ij} = \frac{T_{ij}}{\sum\limits_l T_{il}}\ .
\end{equation}
After all the movements have been completed, interaction among agents sharing the same current location occurs. At this point, we make a mean-field assumption within each patch, so every agent inside a given location makes the same number of contacts. We also assume that the number of contacts is proportional to the density of each patch via a function $f$ which allows for including different ways of weighting the relevance of population density for the number of interactions inside a given patch. Finally, as we want to reflect the mostly commuting nature of human mobility, we force all the individuals to return to their residence and repeat the same process for a new time step (day). 

Under these assumptions, the dynamics is totally characterized by a set of $2\times N$ coupled discrete equations governing the temporal evolution of the fraction of infected and recovered individuals with residence in each patch. In particular, the fraction of infected, $\rho^{I}_i(t+1)$, and recovered individuals, $\rho^{R}_{i}(t+1)$, associated to patch $i$ at time $t+1$, read:
\begin{eqnarray}
\rho^I_i(t+1)&=& (1-\mu) \rho^I_i (t) + (1-\rho^I_i(t)-\rho^R_i(t))\Pi_i (t)\ ,\label{eq:rho}\\
\rho_i^R(t+1)&=& \rho^R_i(t)+ \mu\rho^I_i (t)\ .\label{eq:rhor}
\end{eqnarray}

Eq.~(\ref{eq:rhor}) determines the evolution of recovered patients. As the SIR model assumes that the compartment $R$ constitutes the final epidemiological state, this evolution is just given by the number of infected individuals overcoming the disease. Regarding the evolution of infectious individuals, the r.h.s. of Eq.~(\ref{eq:rho}) corresponds to those infected overcoming the disease and the second one involves contagions of susceptible individuals. In this sense, the probability that a susceptible individual living inside $i$ contracts the disease at time $t$, $\Pi_i (t)$, can be expressed as:
\begin{equation}
\Pi_i (t) = (1-p) P_i (t) + p\sum\limits_{j=1}^N R_{ij} P_j(t)\ .
\label{eq:Pi_i}
\end{equation}
The first term identifies those contagions occurring inside the residence patch whereas the second term contains those taking place inside neighboring areas. Likewise, the probability of contracting the disease inside a given node $i$ at time $t$, $P_i(t)$, is given by:
\begin{equation}
P_i (t) = 1-\left(1-\lambda \frac{I^{eff}_i (t)}{n^{eff}_i}\right)^{f_i}\ , 
\label{eq:P_i}
\end{equation}
where $f_i$ determines the number of contacts per day of individuals inside $i$. Finally, the terms $n^{eff}_i$ and $I^{eff}_i (t)$, which denote the effective population and effective number of infected individuals inside patch $i$ at time $t$ after population movements, read:
\begin{eqnarray}
n^{eff}_i &=& \sum\limits_{j=1}^N n_j\left(\delta_{ij}(1-p) + R_{ji}\right)\ , \\
I^{eff}_i &=& \sum\limits_{j=1}^N n_j \rho^I_j (t)\left(\delta_{ij}(1-p) + R_{ji}\right)\ .
\end{eqnarray}
\subsubsection{Estimating cities' vulnerability}
The former equations offer the possibility of estimating the vulnerability for each city. For this purpose, we study the epidemic threshold defined as the minimum infectivity per contact needed to observe an epidemic outbreak. Therefore, the lower the epidemic threshold is, the easier an epidemic wave propagates, thus reflecting a higher city vulnerability to epidemic outbreaks. To estimate the epidemic threshold, we assume that the disease has reached a stationary equilibrium and that the epidemic size is negligible compared with the population size. Mathematically, this imply that $\vec{\rho}
^I(t+1) = \vec{\rho} (t) = \vec{\epsilon} \ll \vec{1} $. In addition, we neglect the individuals belonging to the compartment $R$ by setting $\vec{\rho}
^R(t+1) = \vec{\rho}^R(t) = \vec{0} $. Both assumptions allow us to linearize Eq. (\ref{eq:P_i}) which now reads:
\begin{equation}
P_i \simeq \lambda f_i\frac{I^{eff}_i}{n^{eff}_i}\ .
\label{eq:linearP}
\end{equation}
After introducing Eqs.(\ref{eq:Pi_i}-\ref{eq:linearP}) into Eq.(\ref{eq:rho}) and taking into account the stationary regime, we obtain:
\begin{equation}
 \frac{\mu}{\lambda}\epsilon_i=\sum\limits_{j=1}^N\underbrace{\left[(1-p)^2\delta_{ij}\frac{f_i}{n_i^{eff}}+ p(1-p)\left(R_{ij}\frac{f_j}{n_j^{eff}}+R_{ji}\frac{f_i}{n_i^{eff}}\right)+p^2\sum\limits_l R_{il}R_{jl}\frac{f_l}{n_l^{eff}}\right]n_j}_{M_{ij}}\epsilon_j\ .
 \label{eq:epsilon}
\end{equation} 
The former expression holds if $\frac{\mu}{\lambda}$ corresponds to an eigenvalue of matrix ${\bf M}$. As our goal is obtaining the minimum $\lambda$ value triggering epidemic outbreaks, the epidemic threshold $\lambda_c$ is given by:
\begin{equation}
\lambda_c = \frac{\mu}{\Lambda_{\text{max}} ({\bf M})}\ ,
\end{equation}
with $\Lambda_{\text{max}} ({\bf M})$ denoting the spectral radius of matrix ${\bf M}$.

\subsection{Function governing contacts}
\label{subsec:lambdaproxy}
At this point, it is necessary to specify the form of the function $f$, determining how the number of contacts that each individual makes inside each patch depends on its density. To shed light on the role of mobility inside each city, we follow a non-parametric approach by assuming that these contacts are linearly proportional to the density inside each patch. This way, the results shown in Figs.~2 \& 4 are obtained by assuming:
\begin{equation}
    f_i = \frac{n^{eff}_i}{a_i}\ ,
\end{equation}
being $a_i$ the area of patch $i$.

If one is interested in isolating the role played by mobility in shaping cities' vulnerability, we must remove the dependence of the overall population density for each case. To do so, we decide to compute the normalized epidemic threshold $\tilde{\lambda}_c$ which is defined as the ratio between the actual epidemic threshold, computed by accounting for human mobility, and the threshold corresponding to a static scenario. Therefore, this quantity reads:
\begin{equation}
    \tilde{\lambda}_c = \frac{\Lambda_{\text{max}} ({\bf M}) (p=0)}{\Lambda_{\text{max}} ({\bf M}) (p=1)}
\end{equation}

\begin{figure}[t!]
    \centering
    \includegraphics[width=0.85\columnwidth]{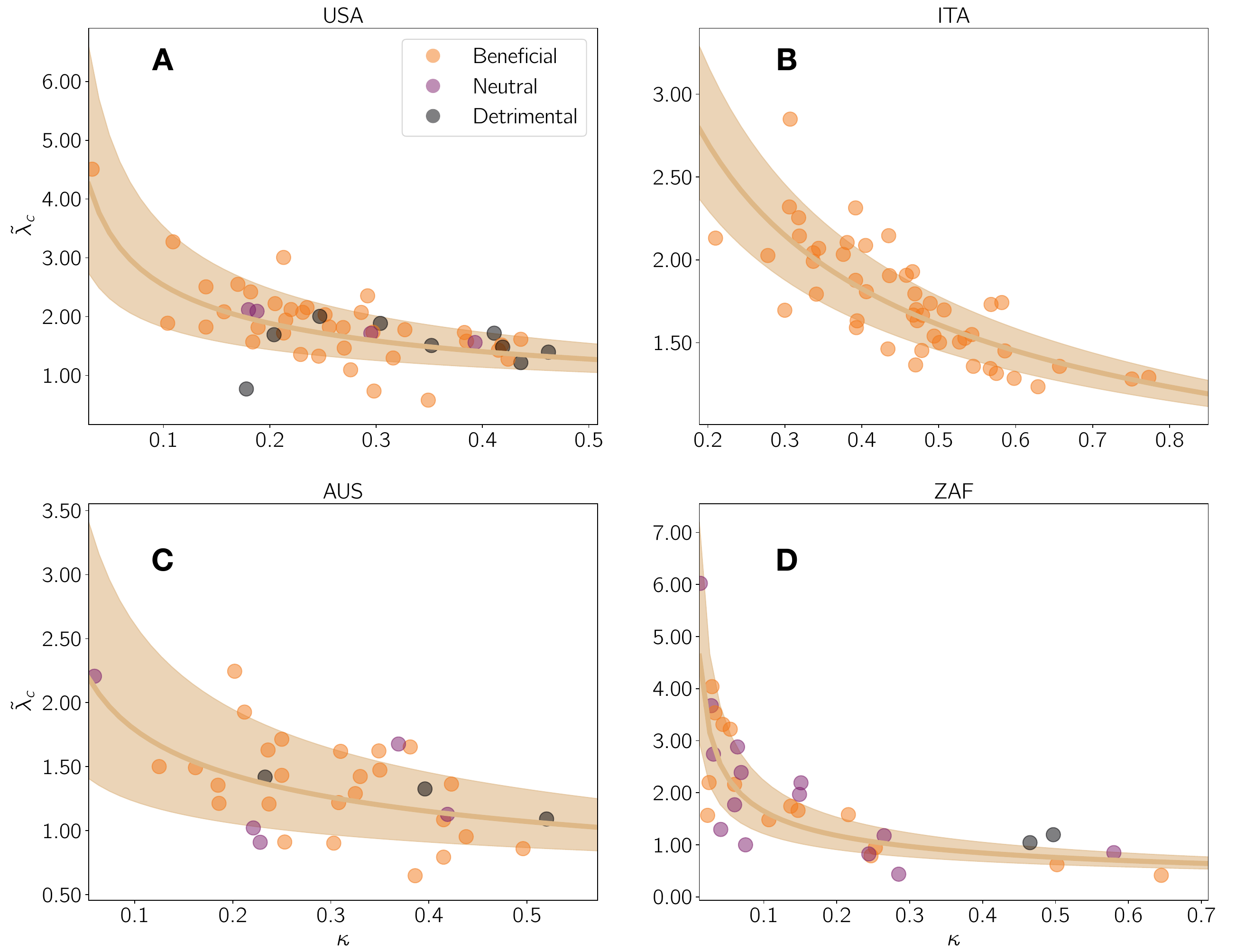}
    \caption{Normalized epidemic threshold versus population density hotspot concentration for the ({\bf A}) United States (Spearman: -.62, Pearson = -.59), ({\bf B}) Italy (Spearman = -.81, Pearson = -.77 ), ({\bf C}) South Africa (Spearman = -.79, Pearson = -.57), and ({\bf D}) Australia (Spearman = -.45, Pearson = -.47) , with each city colored according to the outcome of removing flows connecting hotspots and distribute them evenly among non-hotspots neighboring areas (Fig. 2{\bf B}). Solid line shows the power-law fitting $\tilde{\lambda}_c=A\kappa^{\beta}$ and shadowed regions cover $68\%$ confidence interval.}
    \label{fig:si_hh_thresh}
\end{figure}

Although useful for illustrating the role of mobility in each city, the non-parametric linear relation does not correspond with a realistic scenario due to the large difference in terms of contacts existing among zones with disparate densities. To make a more fair comparison on the expansion of COVID-19 over different cities, we choose a more complex function governing the number of contacts. Following \cite{HU2013125}, we assume that
\begin{equation}
    f_i = 2 - e^{-\xi\frac{n^{eff}_i}{a_i}}\ ,
\end{equation}
which is bounded such that $f_i \in [1,2)$. Here the parameter $\xi$ is estimated by maximizing the correlation among the theoretical and the observed vulnerabilities, yielding $\xi=2\cdot 10^{-5}$ square miles. By including this function, we estimate the epidemic threshold for each city, $\lambda_c$, as:
\begin{equation}
\lambda_c=\frac{\mu}{\Lambda_{\text{max}} ({\bf M}) (p=1)}\ ,
\label{eq:lambdaMOB}
\end{equation}
where we set $\mu=1$ for the sake of simplicity. Note that this parameter does not have any influence on the cities' ranking since it is inherent to the disease features and does not depend on human interactions.

\section{Examining the impact of hotspot concentration at the country level}
In the main text, we reveal that all the cities analyzed here, regardless of their associated country, fall in an universal curve governing the dependence of the normalized epidemic threshold $\tilde{\lambda}_c$ on the flow concentration within hotspots $\kappa$. Moreover, we check that this universal curve is well-represented by a sub-linear power law decay function. Nonetheless, despite the robustness of the results, the spatial resolution of the basic units composing the metapopulations is limited by data availability for each country. Therefore, different spatial resolutions are mixed in Fig. 2 of the main text, which partially hinders the close relation between $\tilde{\lambda}$ and $\kappa_c$.

To characterize the relevance of the flows connecting hotspots for epidemic spreading, we split the universal curve presented in the main text and represent in Fig.~\ref{fig:si_hh_thresh} the individual curves for each of the four countries analyzed here. The high Pearson and Spearman correlation coefficients among $\tilde{\lambda}$ and $\kappa$ along with the goodness of the power law fitting for each individual country strengthen our message about the close connection between the normalized epidemic threshold and the hotspot concentration. The analysis at the country level indicates that, while the function governing the relationship is universal, the parameters are dependent on both the spatial resolution at which the data is available and other differences between cities, not considered in this analysis. 

\setlength{\tabcolsep}{20pt}
\begin{table}[t!]
    \centering
      \caption{Exponent of the power law function $\tilde{\lambda}_c=A\kappa^{\beta}$ governing the dependence of the normalized epidemic threshold on the hotspot concentration.\\}
 \begin{tabular}{c c}
    \toprule
          {\bf Country}& {\bf Decay rate $\beta$}  \\[5pt]
          \midrule
         USA & $-0.43 \pm 0.09$   \\[5pt]
     
         ITA & $-0.57 \pm 0.07$ \\[5pt]

         AUS & $-0.32 \pm 0.1$ \\[5pt]
         
         ZAF & $-0.48 \pm 0.06$ \\[5pt]
         \bottomrule
    \end{tabular}
    \label{tab:countrycorr}
\end{table}
\begin{figure}[t!]
    \centering
    \includegraphics[width=\columnwidth]{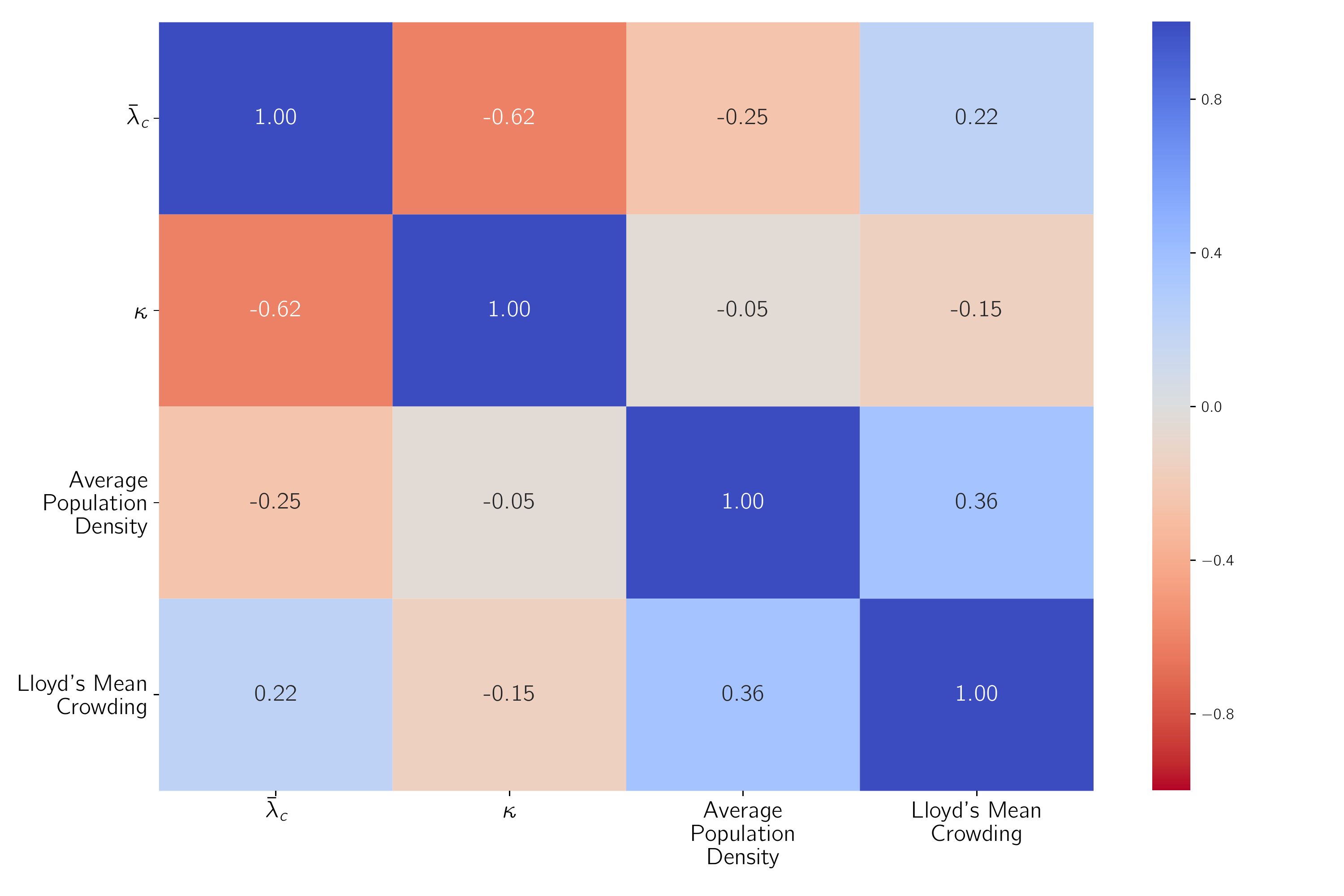}
    \caption{Spearman Correlations between $\kappa$, the average population density in the city, and the Lloyd mean crowding [51] with the normalized epidemic threshold $\tilde{\lambda_c}$}
    \label{fig:pop_corr}
\end{figure}



\section{Connecting Indicators and Population Measures}

We examine the relationship between normalized epidemic threshold $\bar{\lambda}$, population density hotspot concentration $\kappa$ and two other common population measures: average population density and Lloyd' mean crowding for cities in the US in Fig. \ref{fig:pop_corr}.
%
%
%
%
%
%
%
%

\subsection{Average Population Density}
The average was calculated by averaging the population densities $d_{i,k}$ of all $i_{tot}$ zip codes $i$ within each city $k$:

\begin{equation}
    \frac{\sum{d_{i,k}}}{i_{tot}}
\end{equation}

and describes the relative concentration of people within a city.

\subsection{Lloyd Mean Crowding}
Lloyd mean crowding \cite{Scarpino2020} was calculated according to

\begin{equation}
    \frac{\sum{(q_i - 1) q_i}}{\sum{q_i}}
\end{equation}
for patch populations $q_i$. Lloyd mean crowding measures the number of unique contacts possible for members of patches in a city. We note that unlike $\kappa$ and average population density, this measure is independent of patch size and therefore doesn't specifically address proximity.
%

We find that our chosen pair of $\bar{\lambda}$ and $\kappa$ contains the strongest correlations. Of the three population-related metrics $\kappa$, average density, and Lloyd's Mean crowding, $\kappa$ is the only one that contains mobility information, and like average population density, differentiates between patches of different spatial size, supporting that mobility and density are vital inclusions to properly understanding the epidemic situation of a city.

\section{Empirical Epidemic-Indicator Correlations}
\label{sec:exponentialfit}
\begin{figure}[t!]
     \centering
     \includegraphics[width=0.75\columnwidth]{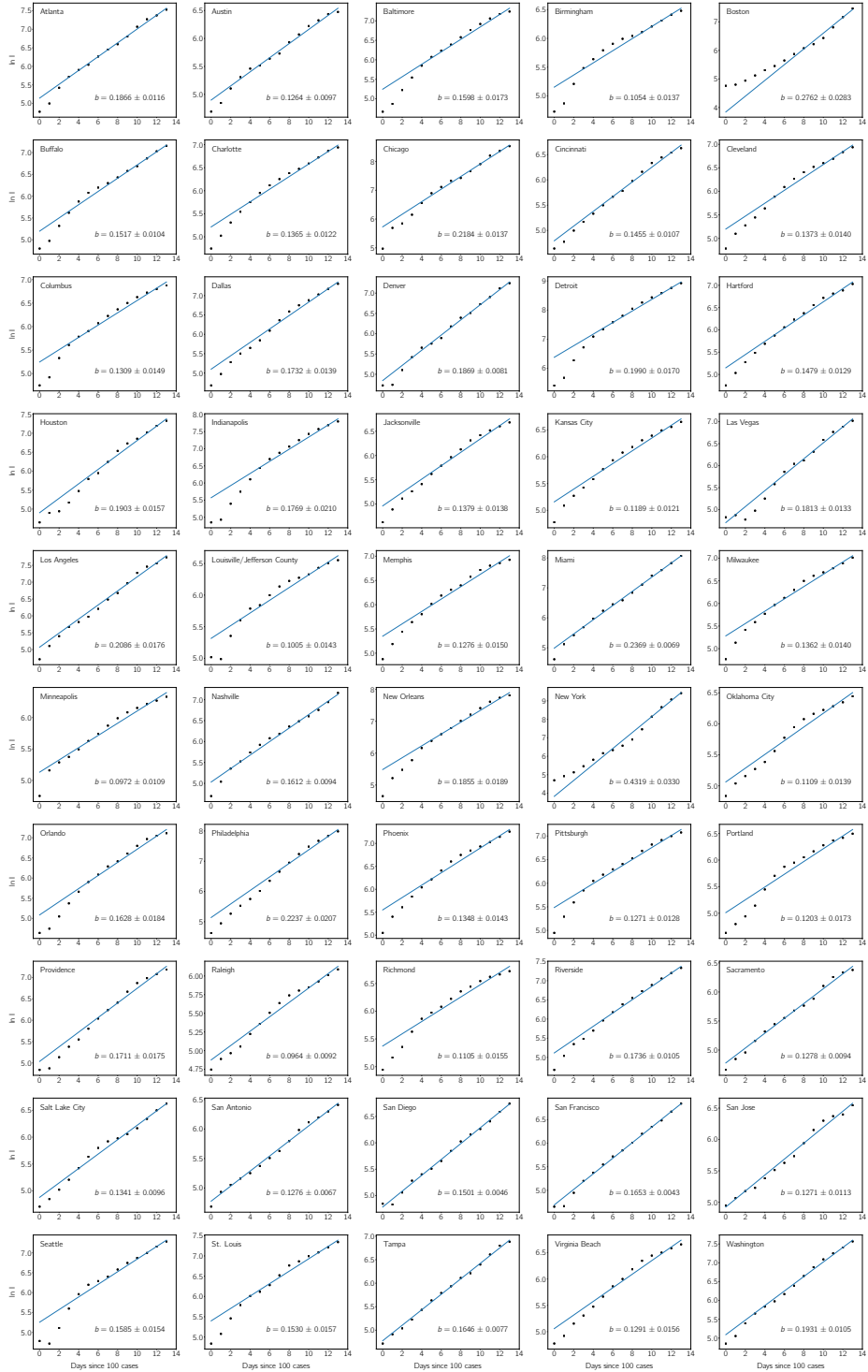}
     \caption{
     Cumulative number of COVID-19 reported 14 days after the first 100 reported cases for the 50 cities in the United States. Dots show real data, smoothed to remove the inherent noisy nature of case reports by using a Savitzky-Golay filter. The line shows the exponential fit of the data via least-squares method to $I(t)=ae^{bt}$.}
     \label{fig:exponential_fit}
 \end{figure}

We quantify the extent to which COVID-19 is able to spread in a US city by examining the timeseries of confirmed cases per county \cite{nytimes,USAfacts}, from January 23 2020 to April 16 2020 (before truncating.) We aggregate this data to the level of CBSAs by summing across each CBSAs component counties. Given the noise in the data (due to collection and reporting artifacts, etc) we perform preprocessing on the curves. We use a Savitzky-Golay Filter \cite{savitzky} to smooth the data by fitting intermediate windows \cite{smooth} with low-order polynomials. We then truncate our data to a window of two weeks after 100 cases were confirmed in each county. This allows our window of observation to capture the regime where COVID-19 awareness encouraged active testing, but before intervention methods influence how the disease propagates within cities. This way, we capture the disease behavior specific to the city structure, and not external suppression. To estimate the vulnerability of each city, we fit the filtered cumulative number of infections to an exponential function
\begin{equation}
    I(t) = ae^{bt}
\end{equation}
and extract the growth rate $b$. Although more sophisticated approaches have been proposed in the literature based on the estimation of the effective reproductive number $R_e$, our procedure is simple but effective to capture the vulnerability of each city at the early stage of the outbreak. The smoothed curves along with the exponential fits are presented in Fig.~\ref{fig:exponential_fit}.
\clearpage
\end{document}